\documentstyle[12pt,prd,aps,epsfig,floats,preprint,eqsecnum]{revtex}
\tightenlines
\newcommand{\sgn}{\mathop{\rm sgn}\nolimits}

\begin{document}

\begin{titlepage}
\preprint{
\vbox{   \hbox{CERN-TH/2001-214}
         \hbox{IFIC-01-43}
         \hbox{hep-ph/0108073}
         }}

\title{Update on Solar and Atmospheric Four-Neutrino Oscillations
\footnote{Expanded version of the proceedings for the talk presented at
the EPS Conference, Budapest, Hungary, July 2001}}
\author{M.~C.~Gonzalez-Garcia$^{1,2,3}$ \thanks{concha@thwgs.cern.ch},
  M.~Maltoni$^{2}$ \thanks{maltoni@ific.uv.es},
  C.~Pe\~na-Garay$^{2}$ \thanks{penya@ific.uv.es} }
\vskip 1cm
\address{
  $^1$ Theory Division, CERN, CH-1211 Geneva 23, Switzerland\\
  $^2$Instituto de F\'{\i}sica Corpuscular,  
  Universitat de  Val\`encia -- C.S.I.C.\\
  Edificio Institutos de Paterna, Apt 22085, 46071 Val\`encia, Spain\\
  $^{3}$ C.N. Yang Institute for Theoretical Physics\\
  State University of New York at Stony Brook\\
  Stony Brook,NY 11794-3840, USA\\}
\maketitle
\begin{abstract}
    In this talk we present the update (including the recent SNO results) 
    of our analysis of the neutrino oscillation solutions of the solar
    and atmospheric neutrino problems in the framework of four-neutrino
    mixing where a sterile neutrino is added to the three standard ones and
    the mass spectra present two separated doublets. 
    Such scenarios allow for
    simultaneous transitions of solar $\nu_e$, as well as atmospheric
    $\nu_\mu$, into active and sterile neutrinos controlled by the additional
    mixing angles $\vartheta_{23}$ and $\vartheta_{24}$, and they contain as
    limiting cases the pure solar $\nu_e$--active and $\nu_e$--sterile
    neutrino oscillations, and the pure atmospheric $\nu_\mu\to\nu_s$ and
    $\nu_\mu\to\nu_\tau$ oscillations, respectively. We evaluate the allowed
    active--sterile admixture in both solar and atmospheric oscillations from
    the combined analysis. Our results show that, although the
    Super-Kamiokande data disfavour both the pure $\nu_\mu\to\nu_s$
    atmospheric channel and, in combination with SNO, the pure 
    $\nu_e\to\nu_s$ solar channel, the result
    from the combined analysis still favours close-to-pure active and
    sterile oscillations and disfavours oscillations into a near-maximal
    active--sterile admixture.
\end{abstract}
\end{titlepage}

\section{Introduction}
\label{intro}

Super-Kamiokande high statistics data~\cite{skcont,skup,skatm00} indicate
that the observed deficit in the $\mu$-like atmospheric events is due to the
neutrinos arriving at the detector at large zenith angles, strongly suggestive
of the $\nu_\mu$ oscillation hypothesis. Similarly, their data on the zenith
angle dependence and recoil energy spectrum of solar
neutrinos~\cite{sksol,sksol00} in combination with the results from
Homestake~\cite{chlorine}, SAGE~\cite{sage}, and GALLEX+GNO~\cite{gallex,gno}
experiments and the recent SNO results~\cite{sno}, have put on a firm 
observational basis the long-standing problem
of solar neutrinos, strongly indicating the need for $\nu_e$ conversions.

In addition to the solar and atmospheric neutrino results from underground
experiments, there is also an indication for neutrino oscillations in the
$\bar\nu_\mu \rightarrow \bar\nu_e$ channel by the LSND
experiment~\cite{lsnd}. All these results can be accommodated in a single
oscillation framework only if there are at least three different scales of
neutrino mass-squared differences. The simplest case in which this condition
is satisfied requires the existence of a fourth light neutrino, which must be
{\it sterile} ({\it i.e.}\ have interactions with standard model particles
much weaker than the SM weak interaction) in order not to affect the invisible
$Z$ decay width, precisely measured at
LEP~\cite{four-models,four-phenomenology,BGG-AB,Barger-variations-98,DGKK-99}.

There are various analyses in the literature that compare the oscillation
channels into active or sterile neutrinos for both
solar~\cite{ourtwo-solar,two-solar,nu2000,bgp} and atmospheric
data~\cite{ourtwo-atmos,two-atmos} performed in the framework of two-neutrino
oscillations (where oscillations occur into either an active or a sterile
state). In particular the Super-Kamiokande collaboration has claimed that the
oscillation into sterile neutrinos is disfavoured for both 
atmospheric~\cite{skatmst} and solar neutrinos~\cite{sksolst}. 
The comparison of the recent SNO CC event rates with the Super-Kamiokande
event rate for solar neutrinos imposes further constraints on 
this channel. However, when
considered in the framework of four-neutrino mixing, oscillations into pure
active or pure sterile states are only limiting cases of the most general
possibility of oscillations into an admixture of active and sterile
neutrinos~\cite{DGKK-99}. Such a more general framework of four-neutrino
mixing has also been analysed in recent studies, which have put constraints on
the active--sterile admixture either in the context of solar~\cite{ourfour} or
atmospheric~\cite{lisi4,lisi4new,yasuda} neutrino oscillations, respectively.

In this talk we present the update of our  {\it combined} analysis of 
the neutrino oscillation
solutions of both the solar and the atmospheric neutrino problem, in the
framework of four-neutrino mixing \cite{globalfour}. 
We include in our analysis the most recent
solar neutrino rates of Homestake~\cite{chlorine}, SAGE~\cite{sage}, GALLEX
and GNO~\cite{gallex,gno}, as well as the recent 1258-day Super-Kamiokande
data sample~\cite{sksol00}, including the recoil electron energy spectrum for
both day and night periods and the recent results from the CC even rates
at SNO~\cite{sno}. As for atmospheric neutrinos we include in our
analysis all the contained events from the latest 79.5 kton-yr
Super-Kamiokande data set~\cite{skatm00}, as well as the upward-going
neutrino-induced muon fluxes from both Super-Kamiokande and the MACRO
detector~\cite{MACRO}. The constraints arising from the relevant reactor 
(mainly from Bugey~\cite{bugey}) and accelerator (CDHSW~\cite{cdhsw} and
CCFR~\cite{ccfr}) experiments are also imposed.

The outline of the paper is the following. In Sec.~\ref{formalism} we
summarize the main expressions for the neutrino oscillation probabilities 
that we use in the analysis of solar and atmospheric neutrino data, 
which take into
account matter effects both in the propagation in the Sun and in the Earth.
Sections~\ref{solar} and~\ref{atmos} contain the results for the analysis of
the four-neutrino oscillation parameters for solar and atmospheric data
respectively. The results for the full combined analysis are described in
Sec.~\ref{combined}, with particular emphasis on the active--sterile
oscillation admixture, which better allows for a simultaneous description of
the solar and atmospheric neutrino data. Our results show that, against what
may be naively expected, although the Super-Kamiokande data disfavour both
the pure $\nu_\mu\to\nu_s$ atmospheric channel and, in combination with 
SNO, the pure $\nu_e\to\nu_s$
solar channel, the result from the combined analysis still favours any of
these close-to-pure active and sterile oscillations and disfavours
oscillations into a near-maximal active--sterile admixture.

\section{four-neutrino Oscillations} 
\label{formalism}

\begin{figure}
    \begin{center} 
        \mbox{\epsfig{file=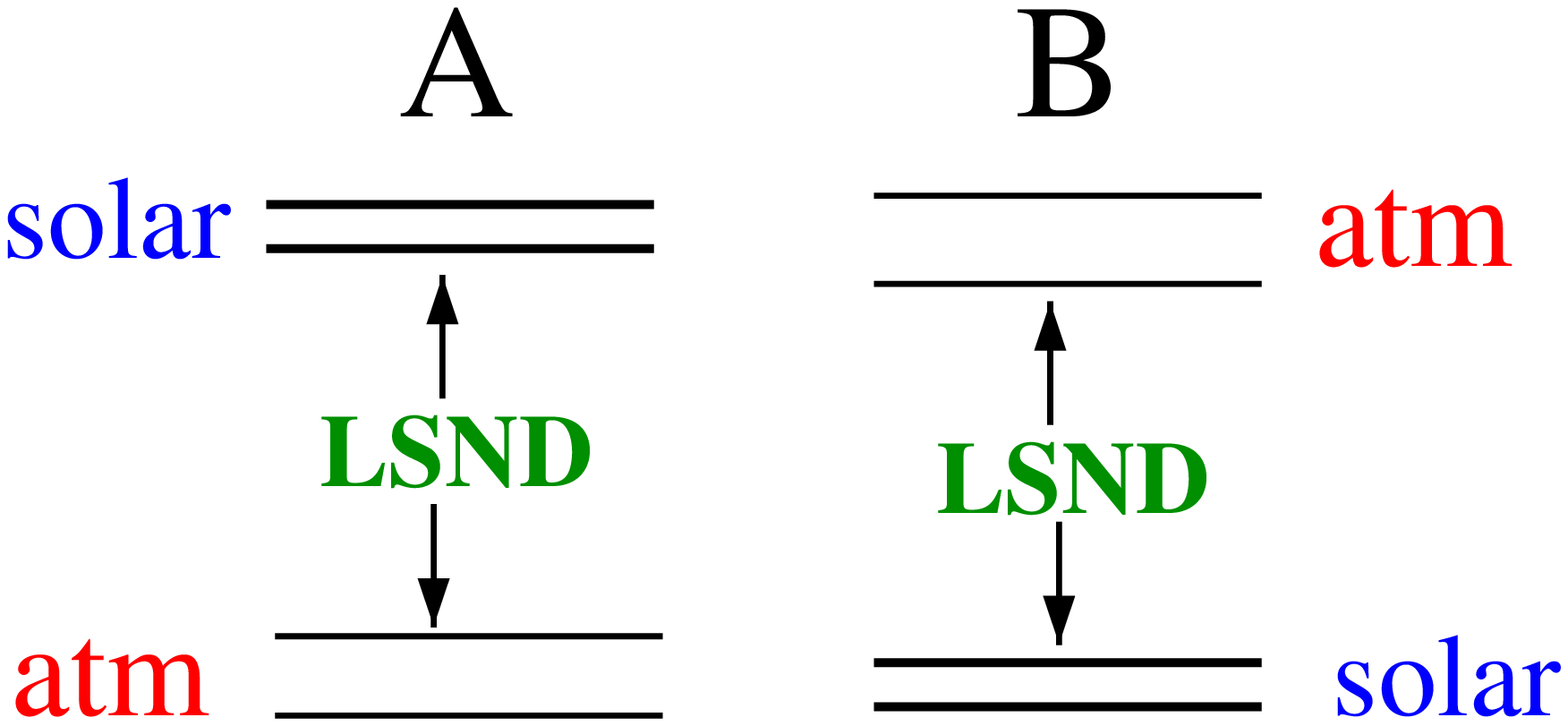,width=0.3\textwidth,angle=-90}} 
    \end{center}
    \caption{$2+2$ mass schemes in four-neutrino mixing favoured from the
      combination of solar, atmospheric and LSND results}.
    \label{fourab}
\end{figure}

There are six possible four-neutrino schemes that can accommodate the results
from solar and atmospheric neutrino experiments as well as the LSND evidence.
They can be divided in two classes: $3+1$ and $2+2$. In the $3+1$ schemes
there is a group of three neutrino masses  separated from an isolated
one by a gap of the order of 1~eV$^2$, which is responsible for the
short-baseline oscillations observed in the LSND experiment. In $2+2$ schemes
there are two pairs of close masses separated by the LSND gap. The $3+1$
schemes are disfavoured by experimental data with respect to the $2+2$
schemes~\cite{BGG-AB,Barger-variations-98}, but they are still marginally
allowed~\cite{3+1}. Therefore we concentrate on the $2+2$ schemes shown in
Fig.~\ref{fourab}. For the phenomenology of neutrino oscillations these two
schemes are equivalent up to the relabelling of the mass eigenstates (or
equivalently of the mixing angles). Thus in what follows we will consider the
scheme B, where the mass spectrum presents the hierarchy:
\begin{equation}
    \Delta m^2_\odot=\Delta m^2_{21} \ll
    \Delta m^2_{\rm atm}=\Delta m^2_{43} \ll
    \Delta m^2_{\rm LSND}=\Delta m^2_{41}\simeq \Delta m^2_{42} \simeq
    \Delta m^2_{31}\simeq\Delta m^2_{32}
\end{equation}
(we use the common notation $\Delta{m}^2_{kj} \equiv m_k^2 - m_j^2$). In this
four-neutrino mixing scheme, the flavour neutrino fields $\nu_{\alpha}$
($\alpha=e,s,\mu,\tau$) are related to the mass eigenstates fields $\nu_{k}$
by
\begin{equation}
    \nu_{\alpha} = \sum_{k=1}^4 U_{\alpha k} \, \nu_{k} \qquad
    (\alpha=e,s,\mu,\tau) \,,
    \label{mixing}
\end{equation}
where $U$ is a $4{\times}4$ unitary mixing matrix. Neglecting possible CP
phases, the matrix $U$ can be written as a product of six rotations, $U_{12}$,
$U_{13}$, $U_{14}$, $U_{23}$, $U_{24}$ and $U_{34}$ where
\begin{equation}
 (U_{ij})_{ab}=\delta_{ab}+
    \left( c_{ij}- 1 \right)
    \left( \delta_{ia} \delta_{ib} + \delta_{ja} \delta_{jb} \right)
    + s_{ij}
    \left( \delta_{ia} \delta_{jb} - \delta_{ja} \delta_{ib} \right)
    \, ,
\end{equation}
where $c_{ij}=\cos{\vartheta_{ij}}$ and $s_{ij}=\sin{\vartheta_{ij}}$. The
order of the product of the matrices corresponds to a specific
parametrization of the mixing matrix $U$.
In order to study oscillations of the solar and atmospheric neutrinos, which
include the matter effects in the Sun and/or in the Earth, it is convenient to
use the following parametrization~\cite{DGKK-99}:
\begin{equation}
    U = U_{24}\,U_{23}\,U_{14}\,U_{13}\,U_{34}\,U_{12} \; .
\end{equation}
This general form can be further simplified by taking into account the
negative results from the reactor experiments, in particular the Bugey
one~\cite{bugey}, which implies that
\begin{equation}
    P_{ee}^{Bugey}=
    1-2\left(|U_{e1}|^2+|U_{e2}|^2\right) \left(1-|U_{e1}|^2-|U_{e2}|^2\right)
    \langle\sin^2\frac{\Delta m^2_{\rm LSND} L}{4 E}\rangle\gtrsim 0.99
\end{equation}
in the range of $\Delta m^2_{41}$ relevant to the LSND experiment. This leads
to an upper bound on the projection of the $\nu_e$ over the 3--4 states:
\begin{equation}
    |U_{e3}|^2+|U_{e4}|^2 = c_{14}^2 s^2_{13}+s_{14}^2 \lesssim 10^{-2}
\end{equation}
so that the two angles $\vartheta_{13}$ and $\vartheta_{14}$ must be small and
their contribution to solar and atmospheric neutrino transitions is
negligible. Therefore, in what follows we will set these two angles to zero,
although it should be kept in mind that at least one of them must be 
non-vanishing in order to account for the LSND observation.

In the approximation $\vartheta_{13} = \vartheta_{14} = 0$ the $U$ matrix
takes the form:
\begin{equation}
    U=
    \left(\begin{array}{cccc} c_{12}& s_{12}& 0& 0\\ 
    - s_{12} c_{23} c_{24}& c_{12} c_{23} c_{24}& s_{23} c_{24} c_{34}
    - s_{24} s_{34}       & s_{23} c_{24} s_{34}+s_{24} c_{34}\\ 
      s_{12} s_{23}       & - c_{12} s_{23}& c_{23} c_{34}& c_{23} s_{34}\\ 
     s_{12} c_{23} s_{24}& - c_{12} c_{23} s_{24} & - s_{23} s_{24} c_{34}
     - c_{24} s_{34}& - s_{23} s_{24} s_{34}+ c_{24} c_{34}
        \end{array}\right)\,.
    \label{Umatrix}
\end{equation}
In the rest of this section we will discuss the consequences of the mixing
structure of Eq.~(\ref{Umatrix}) on the relevant transition probabilities for
solar and atmospheric neutrino oscillations. The transition probabilities that
take into account matter effects have been derived in Ref.~\cite{DGKK-99}.
Some improvement concerning the calculation of the regeneration of solar
$\nu_e$'s in the Earth was presented in Ref.~\cite{ourfour}. Here we summarize
the discussion on the flavour composition of the relevant states for both
solar and atmospheric neutrino oscillations.

\subsection{Solar Neutrinos}

In the scheme here considered, solar neutrino oscillations are generated by
the mass-squared difference between $\nu_2$ and $\nu_1$. It is clear from
Eq.~(\ref{Umatrix}) that the survival of solar $\nu_e$'s mainly depends on the
mixing angle $\vartheta_{12}$, whereas the mixing angles $\vartheta_{23}$ and
$\vartheta_{24}$ determine the relative amount of transitions into sterile
$\nu_s$ or active $\nu_\mu$ and $\nu_\tau$. Let us remind the reader that
$\nu_\mu$ and $\nu_\tau$ cannot be distinguished in solar neutrino
experiments, since their matter potential and their interaction in the
detectors are due only to neutral-current weak interactions and therefore they
are equal. Thus, instead of $\nu_\mu$ and $\nu_\tau$, one can consider the
linear combinations
\begin{equation}
    \left( 
        \begin{array}{l} 
            \nu_a 
            \\ 
            \nu_b 
        \end{array} 
    \right) 
    = 
    \left( 
        \begin{array}{cc} 
            \cos{\vartheta} & \sin{\vartheta} 
            \\ 
            -\sin{\vartheta} & \cos{\vartheta} 
        \end{array} 
    \right) 
    \left( 
        \begin{array}{l} 
            \nu_\mu 
            \\ 
            \nu_\tau 
        \end{array} 
    \right) 
    \,, 
    \label{ab1} 
\end{equation} 
with 
\begin{equation} 
    \cos\vartheta= -\frac{s_{23}}{\sqrt{1-c_{23}^2c_{24}^2}} \, , \qquad
    \sin\vartheta= -\frac{s_{24}c_{23}}{\sqrt{1-c_{23}^2c_{24}^2}} \, .
    \label{ab2} 
\end{equation}
The mixing of $\nu_a$ and $\nu_b$ with $\nu_1$ and $\nu_2$ is given by
\begin{equation}
    U_{a1} 
    = -s_{12} \sqrt{1 - c_{23}^2 c_{24}^2} 
    \,,
    \quad 
    U_{a2} 
    = c_{12} 
    \sqrt{1 - c_{23}^2 c_{24}^2} 
    \,, 
    \quad 
    U_{b1} = U_{b2} = 0 
    \,,
    \label{ab3} 
\end{equation}
so that the solar neutrino oscillations occur only between the states
\begin{equation}
    \nu_e\rightarrow \nu_\alpha \quad
    \mbox{\rm with} \quad
    \nu_\alpha=c_{23} c_{24} \nu_s + \sqrt{1-c_{23}^2 c_{24}^2} \nu_a \; ,
    \label{nusol}
\end{equation}
with mixing angle $\vartheta_{12}$, and
\begin{equation}
    c_{23}^2c_{24}^2 = 1-|U_{a1}|^2-|U_{a2}|^2 = |U_{s1}|^2+|U_{s2}|^2 
\end{equation}
gives the size of the projection of the sterile neutrino onto the state 
in which the solar $\nu_e$ oscillates.

Therefore, solar neutrino oscillations depend only on $\vartheta_{12}$ and the
product $c_{23} c_{24}$. We distinguish the following limiting cases of
Eq.~(\ref{nusol}):
\begin{itemize}
  \item if $c_{23} c_{24} = 0$ then $U_{s1} = U_{s2} = 0$, $U_{a1} = -s_{12}$,
    $U_{a2} = c_{12}$, corresponding to the limit of pure two-generation
    $\nu_e\to\nu_a$ transitions;
  \item if $c_{23} c_{24} = 1$ then $ U_{s1} = s_{12}$, $U_{s2} = c_{12}$ and
    $U_{a1}=U_{a2}=0$ and we have the limit of pure two-generation
    $\nu_e\to\nu_s$ transitions.
\end{itemize}

Since the mixing of $\nu_e$ with $\nu_1$ and $\nu_2$ is described by a
two-generation equation with mixing angle $\vartheta_{12}$, the mixing of
$\nu_s$ with $\nu_1$ and $\nu_2$ is given by the same formula multiplied by
$c_{23} c_{24}$, and the mixing of $\nu_a$ with $\nu_1$ and $\nu_2$ is again
described by the same equation times $\sqrt{ 1 - c_{23}^2 c_{24}^2 }$, it is
clear that, in the general case of simultaneous $\nu_e\to\nu_s$ and
$\nu_e\to\nu_a$ oscillations, the corresponding probabilities are given by
\begin{eqnarray}
    P^{\text{Sun+Earth}}_{\nu_e\to\nu_s}
    & = &
    c^2_{23} c^2_{24}
    \left( 1 - P^{\text{Sun+Earth}}_{\nu_e\to\nu_e} 
    \right) 
    \,,
    \label{Pes} 
    \\ 
    P^{\text{Sun+Earth}}_{\nu_e\to\nu_a} 
    & = &
    \left( 1 - c^2_{23} c^2_{24} 
    \right)
    \left( 1 - P^{\text{Sun+Earth}}_{\nu_e\to\nu_e} 
    \right)
    \,,
    \label{Pea} 
\end{eqnarray}
where $P^{Sun+Earth}_{\nu_e\to\nu_e}$ takes the standard two--neutrino
oscillation form~\cite{nu2000} for $\Delta m^2_{21}$ and $\vartheta_{12}$,
found by numerically solving the evolution equation in the Sun and the Earth
matter with the modified matter potential
\begin{equation}
    A \equiv A_{CC} + c^2_{23}c^2_{24} A_{NC}
    \,,
    \label{A} 
\end{equation}
with $A_{CC} = 2 \sqrt{2} G_F E N_e$ and $A_{NC}=- \sqrt{2} G_F E N_n$. Here
$N_e$ and $N_n$ are, respectively, the number densities of electrons and
neutrons in the medium, $E$ is the neutrino energy and $G_F$ is the Fermi
constant. These expressions satisfy the relation of probability conservation
$P^{\text{Sun+Earth}}_{\nu_e\to\nu_e} + P^{\text{Sun+Earth}}_{\nu_e\to\nu_s} +
P^{\text{Sun+Earth}}_{\nu_e\to\nu_a} = 1$.

The analysis of the solar neutrino data in the four-neutrino mixing schemes
is therefore equivalent to the two-neutrino analysis but taking into account
that the parameter space is now three-dimensional $(\Delta
m^2_{21},\tan^2\vartheta_{12}, c^2_{23} c^2_{24})$. Although originally this
derivation was performed in the framework of the $2+2$
schemes~\cite{DGKK-99,ourfour}, it is equally valid for the $3+1$
ones~\cite{GL}.

Let us finally comment on the range of variation of the mixing parameters.
Choosing the convention that $\Delta m^2_{21}\geq 0$, from Eqs.~(\ref{Pes})
and~(\ref{Pea}) it is clear that for solar neutrino oscillations the full
physical space is covered by choosing the mixing angles in the ranges:
\begin{equation}
    0 \leq \vartheta_{12} \leq \frac{\pi}{2} \;, 
    \qquad
    0 \leq \vartheta_{23} \leq \frac{\pi}{2} \;,
    \qquad
    0 \leq \vartheta_{24} \leq \frac{\pi}{2} \;.
    \label{rangessol}
\end{equation}
Actually, since the dependence on $\vartheta_{23}$ and $\vartheta_{24}$ enters
only through the combination $c_{23}^2 c_{24}^2$, for what concerns the solar
neutrino phenomenology one of these mixing angles could be fixed to zero.
However, as we will see in the next section, they enter independently in the
relevant probabilities for atmospheric neutrinos.

\subsection{Atmospheric Neutrinos}

Atmospheric neutrino oscillations are generated by the mass-squared difference
between $\nu_3$ and $\nu_4$. From Eq.~(\ref{Umatrix}) one sees that the
atmospheric $\nu_e$'s are not affected by the four-neutrino oscillations in
the approximation $\vartheta_{13}=\vartheta_{14}=0$ and neglecting the effect
of $\Delta m^2_{21}$ in the range of atmospheric neutrino energies. 
Conversely, 
the survival probability of atmospheric $\nu_\mu$'s mainly depends on the
mixing angle $\vartheta_{34}$. The mixing angles $\vartheta_{23}$ and
$\vartheta_{24}$ determine the relative amount of $\nu_\mu$ transitions into
sterile $\nu_s$ or active $\nu_\tau$. More explicitly Eq.~(\ref{Umatrix})
implies that the atmospheric neutrino oscillations, {\it i.e.}\ oscillations
with the mass difference $\Delta m^2_{34}$ and mixing angle $\vartheta_{34}$,
occur between the states
\begin{equation}
    \nu_\beta\rightarrow \nu_\gamma 
    \quad
    \mbox{\rm with} 
    \quad
    \nu_\beta = s_{23} c_{24} \nu_s + c_{23}\nu_\mu -s_{23} s_{24} \nu_\tau
    \quad
    \mbox{\rm and}
    \quad
    \nu_\gamma=s_{24} \nu_s +c _{24} \nu_\tau 
    \;.
    \label{nuatm}
\end{equation}
We distinguish the following limiting cases of Eq.~(\ref{nuatm}):
\begin{itemize}
  \item if $c_{23}= 1$ then $U_{\mu 1}=U_{\mu 2}=0$. The atmospheric
    $\nu_\mu=\nu_\beta$ state oscillates into a state $\nu_\gamma=c_{24}
    \nu_\tau +s_{24}\nu_s$. This is the limit studied in
    Refs.~\cite{lisi4,lisi4new}. We will denote this case as ``restricted''. In
    particular:
    \begin{itemize}
      \item in the case $c_{23}=c_{24}=1$ we have $U_{s3}=U_{s4}=0$,
        corresponding to the limit of pure two-generation
        $\nu_\mu\to\nu_\tau$ transitions;
      \item in the case $c_{23}=1$ and $c_{24}=0$ we have $U_{\tau 3} =
        U_{\tau 4} = 0$, corresponding to the limit of pure two-generation
        $\nu_\mu\to\nu_s$ transitions;
    \end{itemize}
  \item if $c_{23}=0$ there are no atmospheric neutrino oscillations as the
    projection of $\nu_\mu$ over the relevant states cancels out
    ($U_{\mu 3}=U_{\mu 4}=0$).
\end{itemize}
Thus the mixing angle $\vartheta_{23}$ determines the size of the projection
of the $\nu_\mu$ over the ``atmospheric'' neutrino oscillating states,
\begin{equation} 
    s^2_{23}=|U_{\mu 1}|^2+ |U_{\mu 2}|^2 =1-|U_{\mu 3}|^2- |U_{\mu 4}|^2.
    \label{muatm}
\end{equation}
As a consequence, one expects $s_{23}$ to be small in order to explain the
atmospheric neutrino deficit and, as we will see in Sec.~\ref{atmos}, this is
the case. Furthermore, the negative results from the CDHSW~\cite{cdhsw} and
CCFR~\cite{ccfr} experiments on searches for $\nu_\mu$-disappearance also
constrain such a projection to be
\begin{equation}
    s^2_{23} = |U_{\mu 1}|^2+ |U_{\mu 2}|^2 \lesssim 0.2
    \label{cdhs} 
\end{equation}
at 90\% CL for $\Delta m^2_{\rm LSND}\gtrsim 0.4$ eV$^2$.

Approximate expressions for the relevant $\nu_\mu$ survival probability
including matter effects in the Earth can be found in Ref.~\cite{DGKK-99}. For
what concerns the work presented here, the relevant probabilities can be
calculated by numerically integrating the evolution equation in the Earth with
a modified matter potential, for example the diagonal piece 
of the potential takes the form
\begin{equation}
    A \equiv (s_{24}^2-s^2_{23}c^2_{24}) A_{NC} \; ,
    \label{Aatm} 
\end{equation}
so that for pure atmospheric $\nu_\mu\rightarrow \nu_\tau$ oscillations
($s_{23}^2 = s_{24}^2 = 0$) $A=0$, while for $\nu_\mu\rightarrow\nu_s$
($s_{23}^2 = 0$, $s_{24}^2 = 1$) $A=A_{NC}$, as expected. It is this
modification of the Earth matter potential that gives the atmospheric
neutrino data the capability to discriminate between the active and sterile
oscillation solution. In particular, higher sensitivity to this potential
effect is achievable for the higher energy part of the atmospheric neutrino
flux, which lead to the upward going muon data. The main effect of the presence
of this potential is that for pure active oscillations the angular
distribution of upgoing muons is expected to be steeper at larger arrival
angles, while a flattening is expected from the matter effects for sterile
oscillations~\cite{lipari}.

From Eqs.~(\ref{nuatm}) and~(\ref{Aatm}) we see that, unlike the case of
solar neutrinos, the angles $\vartheta_{23}$ and $\vartheta_{24}$ enter
independently in the atmospheric oscillations. Thus the analysis of the
atmospheric neutrino data in the four-neutrino mixing schemes is equivalent
to the two-neutrino analysis, but taking into account that the parameter space
is now four-dimensional $(\Delta m^2_{43},\vartheta_{34}, c^2_{23},
c^2_{24})$.

Concerning the range of variation of the mixing angles, the full parameter 
space relevant to atmospheric neutrino oscillation can be covered by choosing
$\vartheta_{23}$ and $\vartheta_{24}$ in the ranges indicated in
Eq.~(\ref{rangessol}), but, in general, the mixing angle $\vartheta_{34}$ must
be allowed to take values in the interval
\begin{equation}
    -\frac{\pi}{2} \leq \vartheta_{34} \leq \frac{\pi}{2} 
    \label{rangesatm}
\end{equation}
(in the convention $\Delta m^2_{43}\geq 0$). In the limiting case
$\vartheta_{23}=0$, there is an additional symmetry in the relevant
probabilities so that the full parameter space can be spanned by
$0\leq\vartheta_{34}\leq\frac{\pi}{2}$, as expected since in this case we have
the effective two--neutrino oscillation $\nu_\mu\rightarrow \nu_\gamma$.

\section{Solar Neutrino Analysis}
\label{solar}

We first describe the results of the analysis of the solar neutrino data in
terms of $\nu_e$ oscillations. We determine the allowed range of oscillation
parameters using the total event rates of the Chlorine~\cite{chlorine},
Gallium~\cite{sage,gallex,gno} as well as the recent SNO result 
on the CC event rate ~\cite{sno}. 
For the Gallium
experiments we have used the weighted average of the results from GALLEX+GNO
and SAGE detectors. We have also included the Super-Kamiokande electron
recoil energy spectrum measured separately during the day and night periods
(corresponding to the 1258-days data sample).
This will be referred to in the following as the day--night spectra data, which
contain $19 + 19$ data bins. The analysis includes the latest standard solar
model fluxes, the BP00 model~\cite{bp00}, with updated distributions 
for neutrino production points and solar matter density.

Details of the statistical analysis applied to the different observables
can be found in  Refs.~\cite{ourtwo-solar,nu2000,bgp}. In particular,
we follow the approach of Ref.~\cite{bgp} and we do not
include here the Super-Kamiokande total rate, since to a large extent
the total rate is represented by the flux in each of the spectral
energy bins.  We define the $\chi^2$ function for the global analysis
as:
\begin{equation}
\chi^2_{\rm sol}=\sum_{i,j=1,41} (R^{th}_i- R^{exp}_i)
\sigma_{G,ij}^{-2} (R^{th}_j- R^{exp}_j) ,
\end{equation}
where $\sigma_{G,ij}^2 = \sigma_{R,ij}^2 + \sigma_{Sp,ij}^2$.  Here
$\sigma_{R,ij}$ is the corresponding $41 \times 41$ error matrix
containing the theoretical as well as the experimental statistical and
systematic uncorrelated errors for the 41 rates while $\sigma_{Sp,ij}$
contains the assumed fully--correlated systematic errors for the
$38\times 38$ submatrix corresponding to the Super-Kamiokande
day--night spectrum data.  We include here the energy independent
systematic error which is usually quoted as part of the systematic
error of the total rate. The error matrix $\sigma_{R,ij}$ includes
important correlations arising from the theoretical errors of the
solar neutrino fluxes, or equivalently of the solar model parameters.
Notice that this statistical treatment is not the same employed
in our pre-SNO analysis~\cite{globalfour} what makes the comparison
not totally straight forward (see Ref.~\cite{bgp} for a 
details on the comparisons).

As discussed in the previous
section, the analysis of the solar neutrino data in these four-neutrino
mixing schemes is equivalent to the two-neutrino analysis but taking into
account that the parameter space is now three-dimensional $(\Delta
m^2_{21},\tan^2\vartheta_{12}, c_{23}^2c_{24}^2)$, so that 
we have a total of 38 degrees of freedom (d.o.f.).

We first present the results of the allowed regions in the three-parameter
space for the global combination of solar observables. Note that, since the
parameter space is three-dimensional, the allowed regions for a given CL are
defined as the set of points satisfying the condition
\begin{equation}
    \chi^2_{\rm sol}(\Delta m_{12}^2,\vartheta_{12},c_{23}^2c_{24}^2)
    -\chi^2_{\rm sol,min}\leq \Delta\chi^2 \mbox{(CL, 3~d.o.f.)} 
\end{equation}
where, for instance, $\Delta\chi^2(\mbox{CL, 3~d.o.f.}) = 6.25$, $7.81$, and
$11.34$ for CL~= 90\%, 95\% and 99\% respectively, and $\chi^2_{\rm sol,min}$
is the global minimum in the three-dimensional space.

\begin{figure}
    \begin{center} 
        \mbox{\epsfig{file=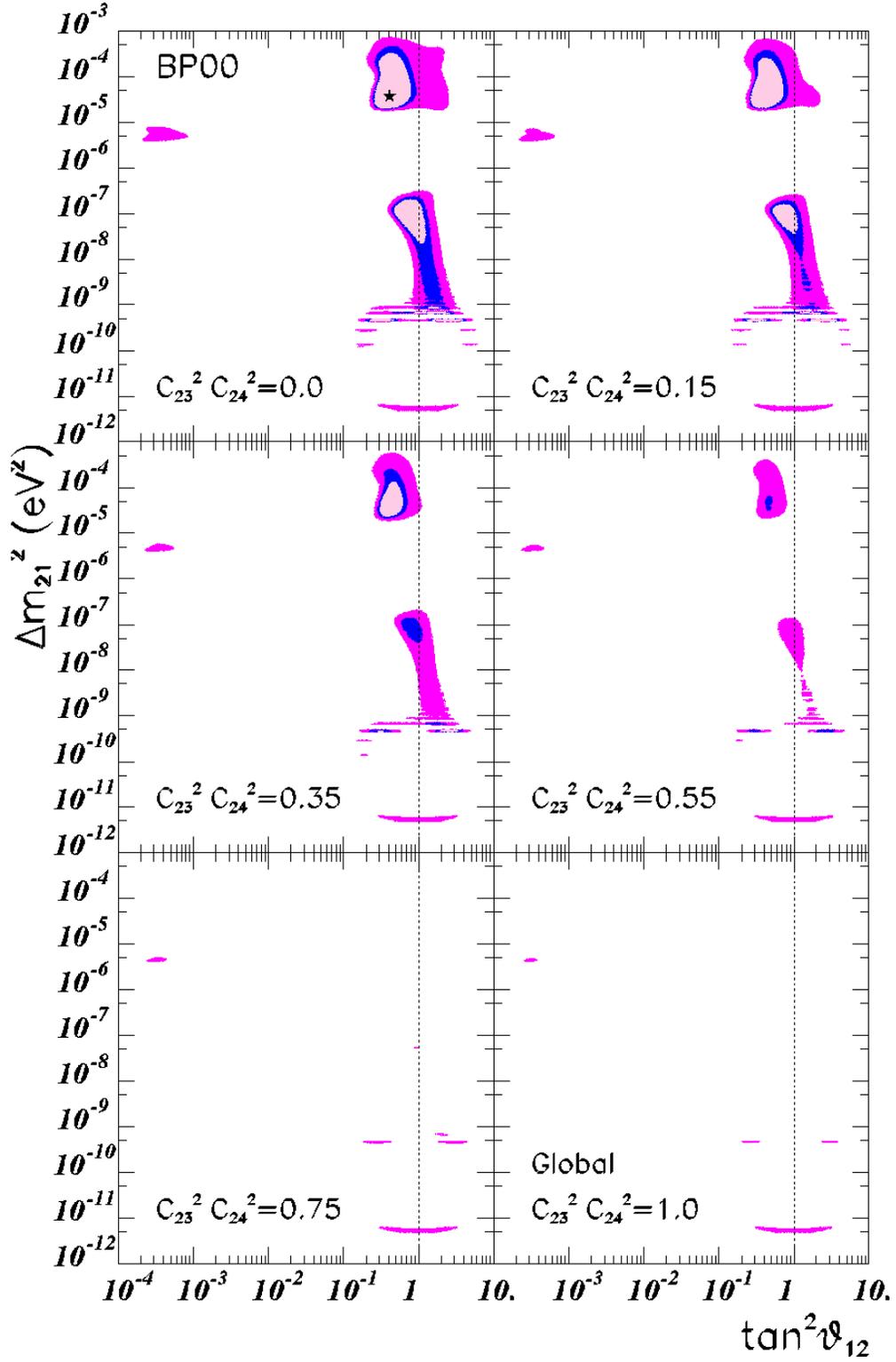,width=0.8\textwidth}}
    \end{center}
    \caption{Results of the global analysis of solar neutrino data for the
      allowed regions in $\Delta{m}^2_{21}$ and $\tan^2 \vartheta_{12}$ for the
      four-neutrino oscillations. The different panels represent sections at
      a given value of the active--sterile admixture $|U_{s1}|^2 + |U_{s2}|^2
      = c_{23}^2 c_{24}^2$ of the three-dimensional allowed regions at 90\%,
      95\% and 99\% CL. The best-fit point in the three-parameter space is
      plotted as a star.}
    \label{solfour}
\end{figure}

In Fig.~\ref{solfour} we plot the sections of such a volume in the plane
($\Delta{m}^2_{21},\tan^2(\vartheta_{12})$) for different values of the
active--sterile admixture $|U_{s1}|^2+|U_{s2}|^2 = c_{23}^2c_{24}^2$. The
global minimum used in the construction of the regions lies in the LMA region
and for pure $\nu_e$--active oscillations
\begin{eqnarray}
    |U_{s1}|^2+|U_{s2}|^2 = c_{23}^2c_{24}^2 & = & 0 \nonumber \\
    \Delta m^2_{21}       & = & 4.1\times 10^{-5}~\mbox{eV}^2 \nonumber \\
    \tan^2\vartheta_{21}  & = & 0.41 \nonumber \\
    \chi^2_{\rm sol,min}  & = & 35.3
\end{eqnarray}
which for 38 d.o.f.\ (3 rates $+$ 38 spectrum points $-$ 3 parameters) 
corresponds to a goodness of the fit (GOF) of
59\%. Notice that 
we have used the Chooz
reactor bound~\cite{chooz} to cutoff the allowed solutions  
for mass values above $\sim$ (7--8) $\times 10^{-4} {\rm eV^2}$. 
The first and last panels shown in Fig.~\ref{solfour} correspond to the
limiting cases of pure active and pure sterile oscillations presented
in Fig.1 of Ref.~\cite{bgp}.

As seen in Fig.~\ref{solfour} the SMA region is still a valid solution for
any value of $c_{23}^2c_{24}^2$ at 99\% CL. This is expected as in the
two-neutrino oscillation picture this solution holds for both pure
$\nu_e$--active and pure $\nu_e$--sterile oscillations. Notice, however, that
the statistical analysis is different: in the two-neutrino picture the pure
$\nu_e$--active and $\nu_e$--sterile cases are analysed separately, whereas in
the four-neutrino picture they are taken into account simultaneously in a
consistent scheme. On the other hand, both the LMA and LOW solutions disappear
for larger values of the active--sterile admixture.

Let us comment here that, unlike the case of atmospheric neutrinos, where
the discrimination between active and sterile oscillations arises from the
difference in the matter potentials, for solar neutrinos the main source of
difference is due to the lack of neutral-current contribution to the water
Cerenkov experiments for the sterile case. Unlike active neutrinos, which lead
to events in the Super-Kamiokande detector by interacting via NC with the
electrons, sterile neutrinos do not contribute to the Super-Kamiokande event
rates. Therefore a larger survival probability for $^8$B neutrinos is needed
to accommodate the measured rate. As a consequence, a larger contribution from
$^8$B neutrinos to the Chlorine and Gallium experiments is expected, so that
the small measured rate in Chlorine can only be accommodated if no $^7$Be
neutrinos are present in the flux. This is only possible in the SMA solution
region, since in the LMA and LOW regions the suppression of $^7$Be neutrinos
is not sufficient. Comparing with our pre-SNO results in Ref.~\cite{globalfour}
we see that, after the inclusion of the SNO result, which observes
a CC event rate for $^8$B neutrinos lower than the total event
rate of Super-Kamiokande, the SMA solution either active or sterile 
is now less likely since it is difficult to fit simultaneously the
Super-Kamiokande flat spectrum and their total event rate and the 
SNO CC measurement.  

Notice also that the SMA region for oscillations into sterile
neutrinos is slightly shifted downwards with respect to the active case. This
is due to the small modification of the neutrino survival probability induced
by the different matter potentials. The matter potential for sterile neutrinos
is smaller than for active neutrinos due to the negative NC contribution,
proportional to the neutron abundance. For this reason the resonant condition
for sterile neutrinos is achieved at lower $\Delta m^2$.

\begin{figure}
    \begin{center}
        \mbox{\epsfig{file=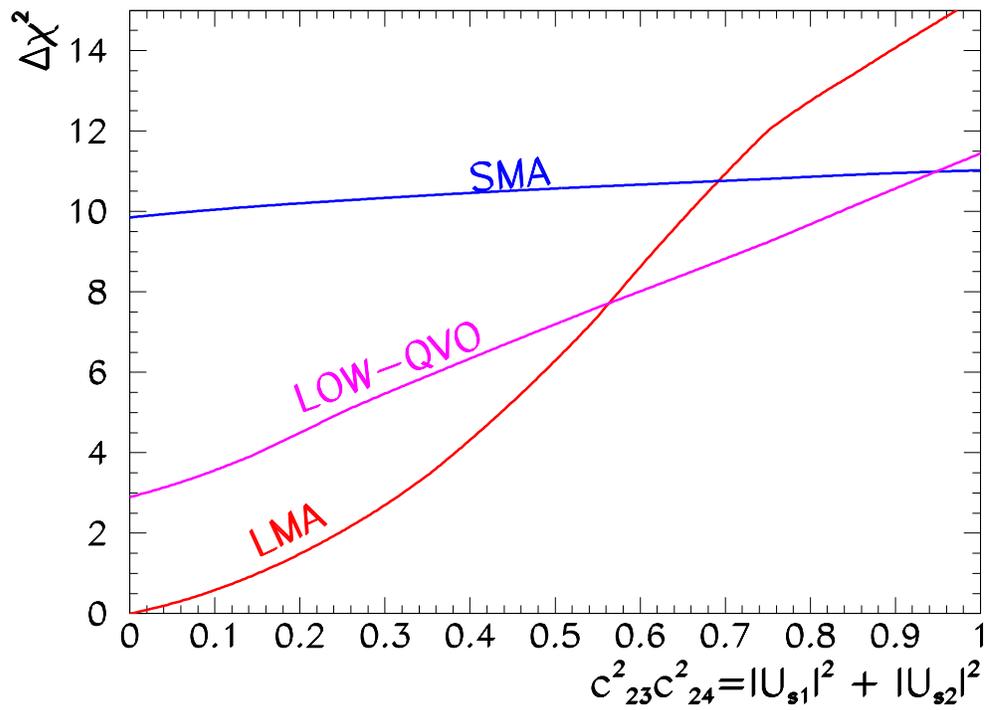,width=0.9\textwidth}} 
    \end{center}
    \caption{Difference $\Delta \chi^2$ between the local minimum of $\chi^2$
      for each solution of the global analysis of solar neutrino data and the
      global minimum in the plane $(\tan^2\vartheta_{12}, \Delta m^2_{21})$ as
      a function of the active--sterile admixture $|U_{s1}|^2 + |U_{s2}|^2 =
      c_{23}^2 c_{24}^2$.}
    \label{chi2sol}
\end{figure}

In Fig.~\ref{chi2sol} we plot the difference $\Delta \chi^2_{\rm sol}$ between
the local minimum of $\chi^2_{\rm sol}$ for each solution and the global
minimum in the three-dimensional space, as a function of the active--sterile
admixture $|U_{s1}|^2+|U_{s2}|^2=c_{23}^2c_{24}^2$. From the figure we find
that:
\begin{itemize}
  \item solar neutrino data favour pure $\nu_e\rightarrow \nu_a$ oscillations
    but sizeable active--sterile admixtures are still allowed;
  \item the dependence of $\Delta \chi^2_{\rm sol}$ on the active--sterile
    admixture is very gentle in the SMA region while it is much more
    pronounced for the LMA solution;
  \item the three-dimensional regions are acceptable at 90\% (99\%) CL for the
    following values of $c_{23}^2c_{24}^2$
    {%
      \catcode`?=\active \def?{\hphantom{0}}
      \begin{eqnarray}
          |U_{s1}|^2+|U_{s2}|^2=c_{23}^2c_{24}^2&<&- \;\; (1.)?? \quad \mbox{\rm for SMA,}\nonumber \\
          |U_{s1}|^2+|U_{s2}|^2=c_{23}^2c_{24}^2&<&0.45 \;\; (0.72) \quad \mbox{\rm for LMA,}\label{limsolar} \\
          |U_{s1}|^2+|U_{s2}|^2=c_{23}^2c_{24}^2&<&0.40 \;\; (0.99) \quad \mbox{\rm for LOW-QVO;} \nonumber
      \end{eqnarray}
      }
  \item at 99\% CL all the three-dimensional  regions are allowed 
    for maximal active--sterile mixing $c_{23}^2c_{24}^2=0.5$. 
  \end{itemize}
There is a subtlety on the meaning of these limiting values of the mixing
$c_{23}^2c_{24}^2$. The values quoted in Eq.~(\ref{limsolar}) correspond to
the mixings at which the corresponding three-dimensional region disappears at
a given CL. Strictly speaking this is not the same as the allowed range of
$c_{23}^2c_{24}^2$ at a given CL when the parameters $(\tan^2\vartheta_{12},
\Delta m^2_{21})$ are left free to vary, either in the full parameter space or
within a given solution. Such limits would be obtained by imposing the
condition $\Delta\chi^2_{\rm sol}=\chi^2_{\rm sol,min\;12}
(c_{23}^2c_{24}^2)-\chi^2_{\rm sol,min}\leq \Delta\chi^2\mbox{(CL, 1~d.o.f.)}$
where $\chi^2_{\rm sol,min \; 12}(c_{23}^2c_{24}^2)$ is minimized in the
parameter space 1--2 (or in a given region of this parameter space) and
$\chi^2_{\rm sol,min}$ is minimized in the same region of the space 1--2 and
in the mixing $c_{23}^2c_{24}^2$.

\section{Atmospheric Neutrino Analysis}
\label{atmos}

In our statistical analysis of the atmospheric neutrino events we use all the
samples of Super-Kamiokande data: $e$-like and $\mu$-like samples of
sub- and multi-GeV~\cite{skatm00} data, each given as a 5-bin zenith-angle
distribution\footnote{Note that for convenience and maximal statistical
significance we prefer to bin the Super-Kamiokande contained event data in 5,
instead of 10 bins.}, and upgoing muon data including the stopping (5 bins in
zenith angle) and through-going (10 angular bins) muon fluxes. We have also
included the latest MACRO~\cite{MACRO} upgoing muon samples, with 10 angular
bins, which is also sensitive to the active--sterile admixture. So we have a
total of 45 independent inputs.

For details on the statistical analysis applied to the different observables, 
we refer to Refs.~\cite{ourtwo-atmos,our3}. As discussed in the previous
section the analysis of the atmospheric neutrino data in these four-neutrino
mixing schemes is equivalent to the two-neutrino analysis, but taking into
account that the parameter space is now four-dimensional $(\Delta
m^2_{43},\vartheta_{34}, c_{23}^2, c_{24}^2)$. Alternatively we present the
results in the equivalent parameter space $(\Delta m^2_{43},\vartheta_{34},
|U_{\mu 1}|^2 + |U_{\mu 2}|^2, |U_{s1}|^2 + |U_{s2}|^2)$.

We first present the results of the allowed regions in the four-parameter
space for the global combination of atmospheric observables. Notice that since
the parameter space is four-dimensional the allowed regions for a given CL are
defined as the set of points satisfying the condition
\begin{equation}
    \chi^2_{\rm atm}(\Delta m_{43}^2,\vartheta_{34},c_{23}^2,c_{24}^2)
    -\chi^2_{\rm atm,min}\leq \Delta\chi^2(\mbox{CL, 4~d.o.f.})
\end{equation}
where $\Delta\chi^2(\mbox{CL, 4~d.o.f.}) = 7.78$, $9.49$, and $13.3$ for CL~=
90\%, 95\% and 99\%, respectively, and $\chi^2_{\rm atm,min}$ is the global
minimum in the four-dimensional space.

\begin{figure}
    \begin{center} 
        \mbox{\epsfig{file=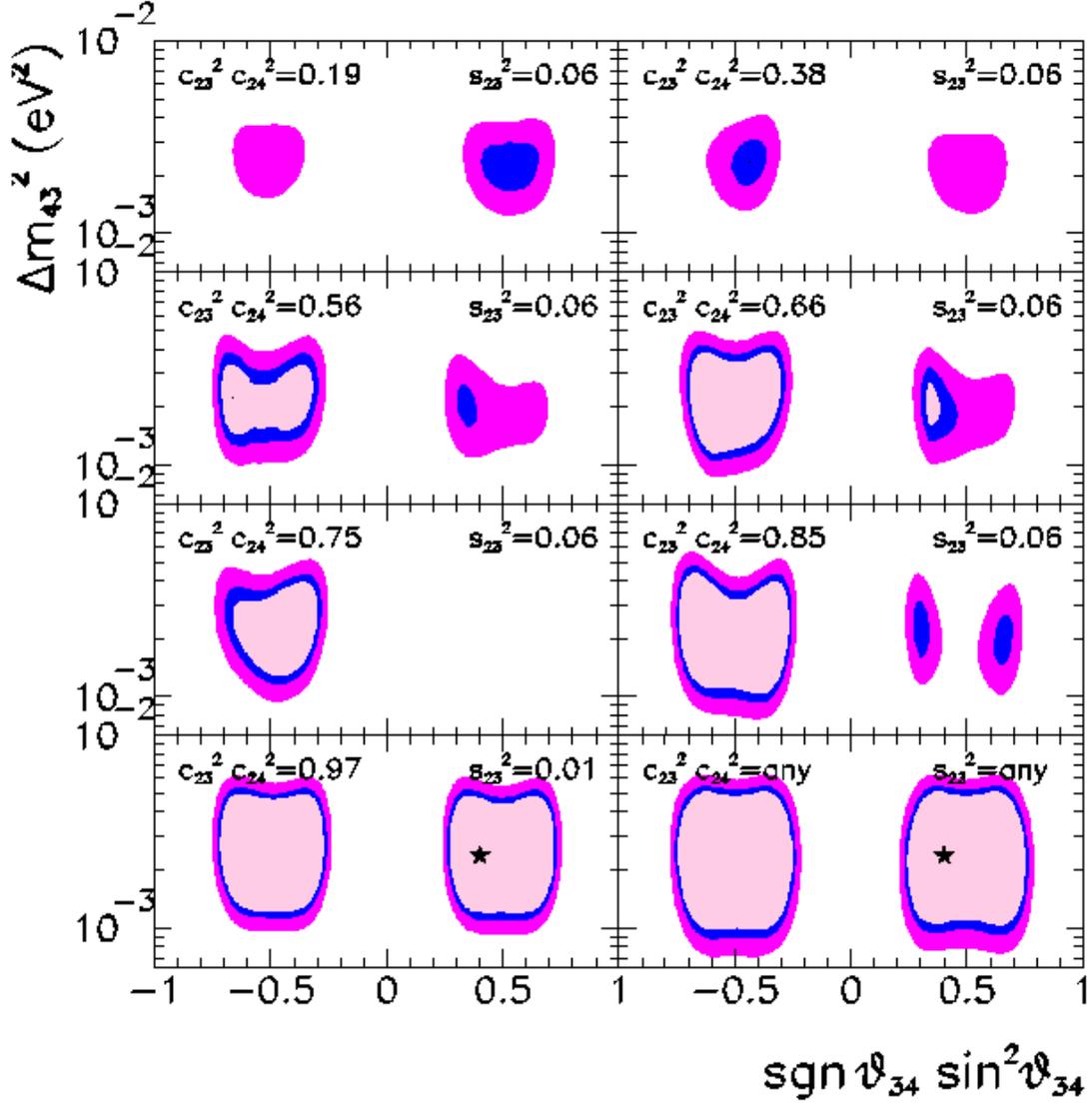,width=0.9\textwidth}} 
    \end{center}
    \caption{Results of the analysis of atmospheric neutrino data for the
      allowed regions in $\Delta{m}^2_{43}$ and $\vartheta_{34}$ for the
      four-neutrino oscillations. The different panels represent sections at
      given values of the $\nu_\mu$ projection $|U_{\mu 1}|^2 + |U_{\mu 2}|^2
      = s_{23}^2$ and the active--sterile admixture $|U_{s1}|^2 + |U_{s2}|^2 =
      c_{23}^2c_{24}^2$ of the four-dimensional allowed regions at 90\%, 95\%
      and 99\% CL. The best-fit point in the four-parameter space is plotted
      as a star. The last panel corresponds to the case in which $\chi^2$ has
      also been minimized with respect to  $s_{23}^2$ and
      $c_{23}^2c_{24}^2$.}
    \label{atmfour}
\end{figure}

In Fig.~\ref{atmfour} we plot the sections of such a volume in the plane
($\Delta{m}^2_{43},\sin^2(\vartheta_{34})$) for different values of the
mixings $\vartheta_{23}$ and $\vartheta_{24}$, which we parametrize by values
of the projections $|U_{\mu 1}|^2+|U_{\mu 2}|^2$ and $|U_{s1}|^2+|U_{s2}|^2$.
As discussed in Sec.~\ref{formalism} for arbitrary values of $c_{23}^2$ and
$c_{24}^2$ the full parameter space is covered for $\vartheta_{34}$ in the
range given in Eq.~(\ref{rangesatm}). In Fig.~\ref{atmfour} we display this
full parameter space by showing the region in ($\Delta{m}^2_{43},
\sgn(\vartheta_{34}) \sin^2(\vartheta_{34})$) with ``positive''
$\sin^2(\vartheta_{34})$ for $0 < \vartheta_{34} \leq \frac{\pi}{2}$ and
``negative'' $\sin^2(\vartheta_{34})$ for $-\frac{\pi}{2} \leq \vartheta_{34}
< 0$. As seen from the figure the parameter space is composed of two separated
regions, each of them around maximal mixing $\vartheta_{34} =
\pm\frac{\pi}{4}$.

The global minimum used in the construction of the regions lies almost in the
pure atmospheric $\nu_\mu$--$\nu_\tau$ oscillations. At the best=fit point
\begin{eqnarray}
    |U_{s1}|^2+|U_{s2}|^2 = c^2_{23} c^2_{24} & = & 0.97 \nonumber \\
    |U_{\mu 1}|^2+|U_{\mu 2}|^2 = s_{23}^2    & = & 0.01 \nonumber \\
    \Delta m^2_{43}      & = & 2.4\times 10^{-3}~\mbox{eV}^2 \nonumber \\
    \vartheta_{34}       & = & 39^\circ \nonumber \\
    \chi^2_{\rm atm,min} & = & 29.0
    \label{minatm}
\end{eqnarray}
which for $45-4=41$ d.o.f.\ corresponds to a GOF of 92\%. A careful reader may
find this GOF surprisingly high. Let us comment here that there has been a
reduction of the $\chi^2$ with respect to previous
analyses~\cite{ourtwo-atmos,our3} due to a better agreement of the 79.5 SK
electron distribution with the non-oscillation expectation (the same effect is
found in Ref.~\cite{lisi4new}). Since the scenarios discussed here do not
affect the atmospheric $\nu_e$ predictions, this implies an overall
improvement on the GOF of any of these four-neutrino oscillation schemes.

From Fig.~\ref{atmfour} we see that the region becomes considerably smaller
for increasing values of the mixing angle $\vartheta_{23}$, which determines
the size of the projection of the $\nu_\mu$ over the ``atmospheric'' neutrino
oscillating states, and for increasing values of the mixing angle
$\vartheta_{24}$, which determines the active--sterile admixture in which the
``almost--$\nu_\mu$'' oscillates. Therefore from the analysis of the
atmospheric neutrino data we obtain an upper bound on both mixings, which, in
particular, implies a lower bound on the combination $c_{23}^2 c_{24}^2 =
|U_{s1}|^2+|U_{s2}|^2$ limited from above by the solar neutrino data. To
quantify these bounds we display in Fig.~\ref{chi2atm}a the allowed region in
the parameter space $(s^2_{23} = |U_{\mu 1}|^2+|U_{\mu 2}|^2, c_{23}^2
c_{24}^2 = |U_{s1}|^2 + |U_{s2}|^2)$. Following the discussion below
Eq.~(\ref{limsolar}) we obtain the allowed two-dimensional region by imposing
the condition
\begin{equation}
    \Delta\chi^2_{\rm atm} = 
    \chi^2_{\rm atm,min \; 34}(\vartheta_{23},\vartheta_{24})
    - \chi^2_{\rm min} \leq \Delta\chi^2(\mbox{CL, 2~d.o.f.})\; ,
    \label{c2dim}
\end{equation}
where $\chi^2_{\rm atm,min \; 34}(\vartheta_{23},\vartheta_{24})$ is minimized
with respect to $\Delta m^2_{43}$ and $\vartheta_{34}$, and $\chi^2_{\rm
atm,min}$ is the global minimum in the four-dimensional parameter space. The
diagonal cut in Fig.~\ref{chi2atm}a gives the unitarity bound
\begin{equation}
    |U_{\mu 1}|^2 + |U_{\mu 2}|^2 + |U_{s 1}|^2 + |U_{s 2}|^2 =
    |U_{\tau 3}|^2 + |U_{\tau 4}|^2 \leq 1\; ,
\end{equation}
where we have used that in Eq.~(\ref{Umatrix}) $U_{e3}=U_{e4}=0$. As seen from
Fig.~\ref{chi2atm}a, the analysis of the atmospheric data imposes a severe
bound on the $\nu_\mu$ projection. For instance, the two-dimensional
region extends only to 
\begin{equation}
    s^2_{23} = |U_{\mu 1}|^2 + |U_{\mu 2}|^2\lesssim 0.12 \; (0.16)
\end{equation}
at 90\% (99\%) CL, which, as previously mentioned (see Eq.~(\ref{cdhs})), is
perfectly consistent with the bounds from the $\nu_\mu$ disappearance
accelerator experiments CDHDW~\cite{cdhsw} and CCFR~\cite{ccfr} given in
Eq.~(\ref{cdhs}).

\begin{figure}
    \begin{center} 
        \mbox{\epsfig{file=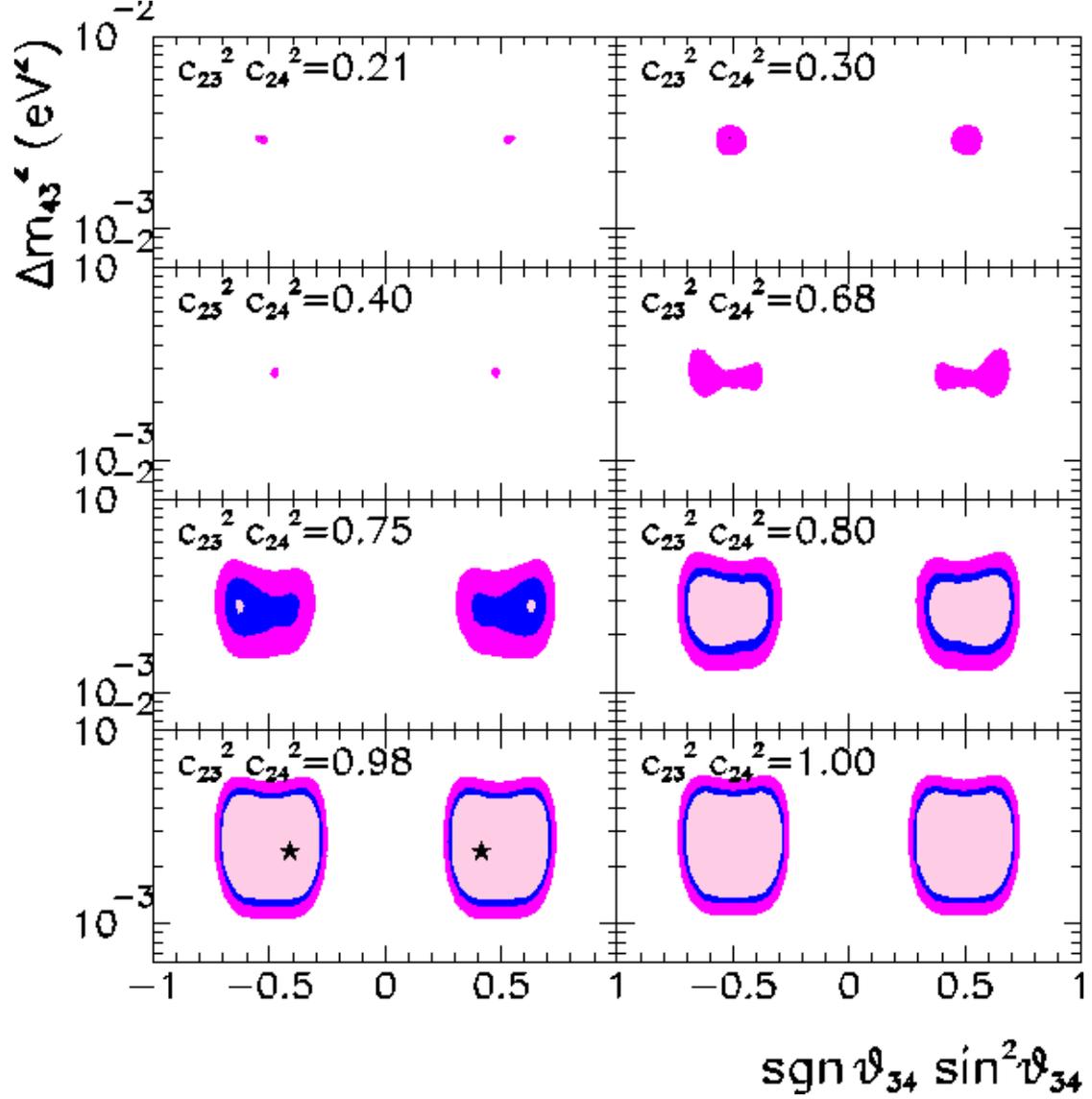,width=0.9\textwidth}} 
    \end{center}
    \caption{Results of the analysis of atmospheric neutrino data for the
      allowed regions in $\Delta{m}^2_{43}$ and $\vartheta_{34}$ for the
      four-neutrino oscillations in the ``restricted'' case of $\nu_\mu$
      projection $|U_{\mu 1}|^2 + |U_{\mu 2}|^2 = s_{23}^2 = 0$. The different
      panels represent sections at given values of the active--sterile
      admixture $|U_{s1}|^2 + |U_{s2}|^2 = c_{24}^2$ of the
      three-dimensional allowed regions at 90\%, 95\% and 99\% CL. The
      best-fit point in the three-parameter space is plotted as a star.}
    \label{atmfourr}
\end{figure}

\begin{figure}
    \begin{center}
        \mbox{\epsfig{file=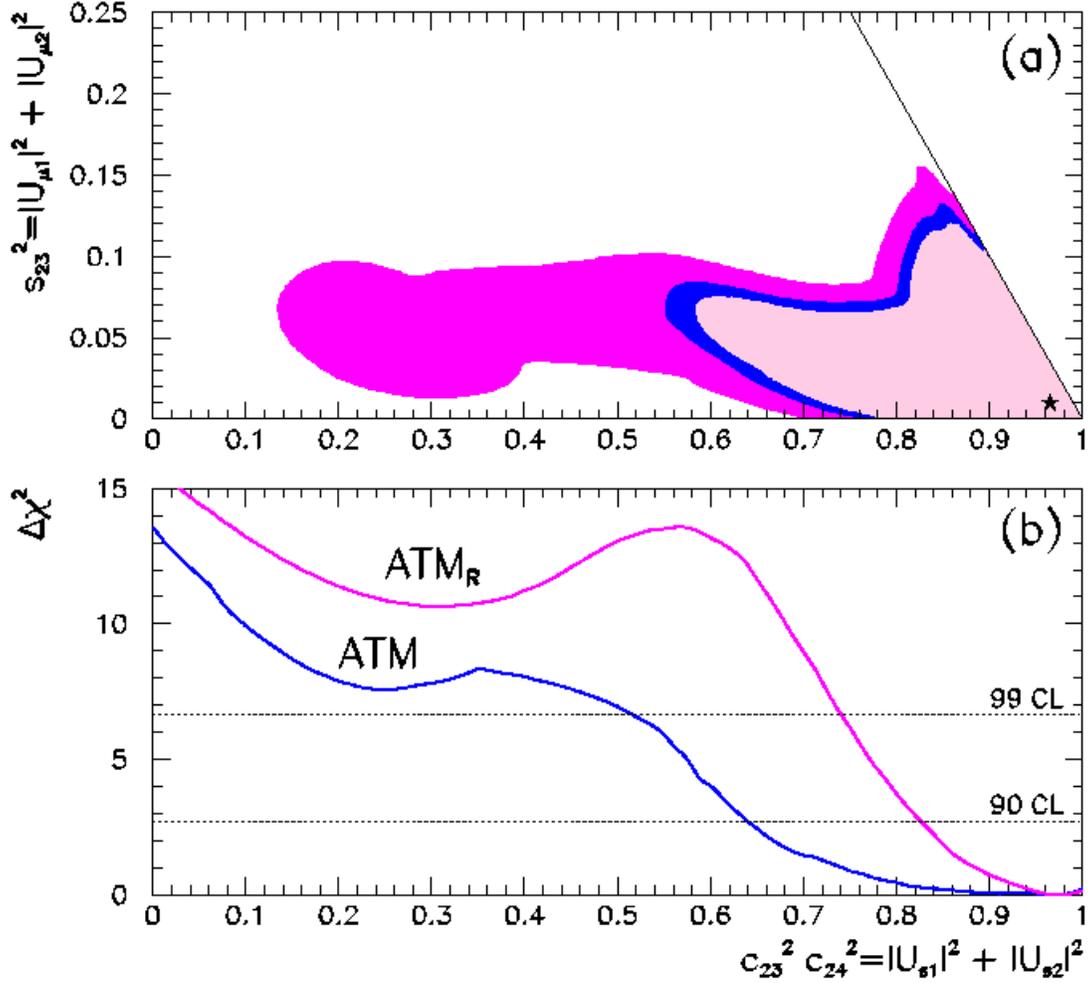,width=0.9\textwidth}}
    \end{center}
    \caption{(a) 90\%, 95\% and 99\% CL allowed regions for the $\nu_\mu$
      projection $|U_{\mu 1}|^2+|U_{\mu 2}|^2 = s_{23}^2$ and the
      active--sterile admixture $|U_{s1}|^2 + |U_{s2}|^2 = c_{23}^2c_{24}^2$
      from the analysis of the atmospheric data (see text for details). (b)
      $\Delta \chi^2$ as a function of the active--sterile admixture
      $|U_{s1}|^2 + |U_{s2}|^2 = c_{23}^2 c_{24}^2$. In the lower curve
      $\chi^2$ has been minimized with respect to $\vartheta_{34}$, $\Delta
      m^2_{43}$ and the $\nu_\mu$ projection $|U_{\mu 1}|^2 + |U_{\mu 2}|^2 =
      s_{23}^2$. The upper curve corresponds to the ``restricted'' case
      $|U_{\mu 1}|^2 + |U_{\mu 2}|^2 = s_{23}^2 = 0$, for which $\chi^2$ has
      been minimized only with respect to $\vartheta_{34}$ and $\Delta
      m^2_{43}$.}
    \label{chi2atm}
\end{figure}

In order to determine the impact of the small but possible deviation from zero
of $s^2_{23}$ we also study the ``restricted'' case of
Refs.\cite{lisi4,lisi4new} in which we impose $s^2_{23} = |U_{\mu 1}|^2 +
|U_{\mu 2}|^2 = 0$. In this case the parameter space for atmospheric neutrino
oscillation is three-dimensional and the allowed regions for a given CL are
defined as the set of points satisfying the condition
\begin{equation}
    \chi^2_{\rm atm}(\Delta m_{43}^2, \vartheta_{34},c_{24}^2)
    - \chi^2_{\rm atm,min}\leq \Delta\chi^2(\mbox{CL, 3~d.o.f.}) \;.
\end{equation}
In Fig.~\ref{atmfourr} we plot the sections of this volume in the plane
$(\Delta{m}^2_{43},\sin^2(\vartheta_{34}))$ for different values of
$c_{24}^2$. As discussed in Sec.~\ref{formalism} for the ``restricted'' case
the full parameter space is covered with $0 \leq \vartheta_{34} \leq
\frac{\pi}{2}$. This is graphically displayed in Fig.~\ref{atmfourr}, where
the symmetry $\vartheta_{34} \to -\vartheta_{34}$ is evident.

The constraints on the active--sterile admixture in the two cases are displayed
in Fig.~\ref{chi2atm}b, where we plot the dependence of $\Delta\chi^2_{atm}$
on the active--sterile admixture $c^2_{23}c^2_{24} = |U_{s1}|^2+|U_{s2}|^2$,
both in the general case ({\it i.e.}\ when $\chi^2_{\rm atm}$ is minimized
with respect to all other parameters in the problem) and in the
``restricted'' case (for which the $\nu_\mu$ projection has been fixed
$|U_{\mu 1}|^2+|U_{\mu 2}|^2=0$). The horizontal lines are the values of
$\Delta\chi^2(\mbox{CL, 1~d.o.f.})$ for CL~= 90\% and 99\%. From this figure
we obtain the 90\% (99\%) CL lower bounds on $c^2_{23}c^2_{24}$ from the
analysis of the atmospheric neutrino data:
\begin{eqnarray}
    c^2_{23}c^2_{24}=|U_{s 1}|^2+|U_{s 2}|^2 & > & 0.64\; (0.52) \\
    c^2_{24}=|U_{s 1}|^2+|U_{s 2}|^2         & > & 0.83\; (0.74) \quad
    \mbox{\rm for the restricted case.}
\end{eqnarray}
As seen in the figure $\Delta\chi^2_{\rm atm}$ has a monotonically growing
behaviour as $|U_{s 1}|^2 + |U_{s 2}|^2$ decreases for $|U_{s 1}|^2 + |U_{s
2}|^2 \gtrsim 0.4$--0.6, while the dependence flattens below that value
with the presence of a shallow secondary minimum around $|U_{s 1}|^2 + |U_{s
2}|^2\sim 0.25$--0.3. We have traced the presence of this secondary
minimum to the combined effect of the behaviour of $\chi^2$ for the different
sets of upgoing muon data. For all sets of upgoing muon data the best-fit is
very close to the global minimum in Eq.~(\ref{minatm}). MACRO data at the
moment lead to a monotonically growing $\chi^2$, as the sterile admixture
grows while Super-Kamiokande upgoing data on both stopping and through-going
muons presents a flattening of their $\chi^2$ for $|U_{s 1}|^2 + |U_{s 2}|^2
\le 0.5$--0.6. This loss of sensitivity of Super-Kamiokande at larger
sterile admixtures arises from the fact that their latest upgoing muon data
presents a flattening of the muon suppression in the angular bins for arrival
directions around $-0.7 \leq \cos\theta \leq -0.4$, which is compatible with
large sterile admixtures. The same effect has been found in
Ref.~\cite{lisi4new}. When combining the $\chi^2$ for MACRO and
Super-Kamiokande data samples, and taking into account the strong correlation
between their respective theoretical errors, we find the appearance of this
secondary shallow minimum.

From the figure we see that oscillations into a pure sterile state are
excluded at 99.97\% CL as compared to pure active oscillations. One must
notice, however, that taken as an independent scenario, $\nu_\mu \to \nu_s$
oscillations give a $\chi^2_{min} = 46$ (for $45-2$ d.o.f.), which implies a
GOF of 35\% and therefore cannot be ruled out on the basis of its absolute
quality.

In summary, we see that the analysis of the atmospheric data implies that:
\begin{itemize}
  \item one of the two-neutrino states in the 3--4 oscillating pair must be
    a close-to-pure $\nu_\mu$;
  \item this close-to-pure $\nu_\mu$  oscillates into a state which is
    preferably composed by $\nu_\tau$, although a 48\% admixture of sterile in
    this state is still allowed at 99\% CL;
  \item when imposing the ``restricted'' condition that one of the states is a
    pure $\nu_\mu$, the bound on the sterile admixture is tightened
    approximately by a factor 2 (from 48\% to 26\% at 99\% CL).
\end{itemize}

\section{Combined Analysis: Conclusions}
\label{combined}

From the previous sections we have learned that the analysis of the solar data
favours the scenario in which the solar oscillations in the plane 1--2 are
$\nu_e$ oscillations into an active neutrino, and from that analysis we found
an upper limit on the projection of the $\nu_s$ on the 1--2 states. On the
other hand, the atmospheric neutrino analysis prefers the oscillations of
the 3--4 states to occur between a close-to-pure $\nu_\mu$ and an active
($\nu_\tau$) neutrino, thus giving an upper bound on the projection of the
$\nu_s$ over the 3--4 states, or equivalently a lower bound on its projection
over the 1--2 states. The open question is then what the best scenario is for
the active--sterile admixture once these two bounds are put together.

\begin{figure}
    \begin{center}
        \mbox{\epsfig{file=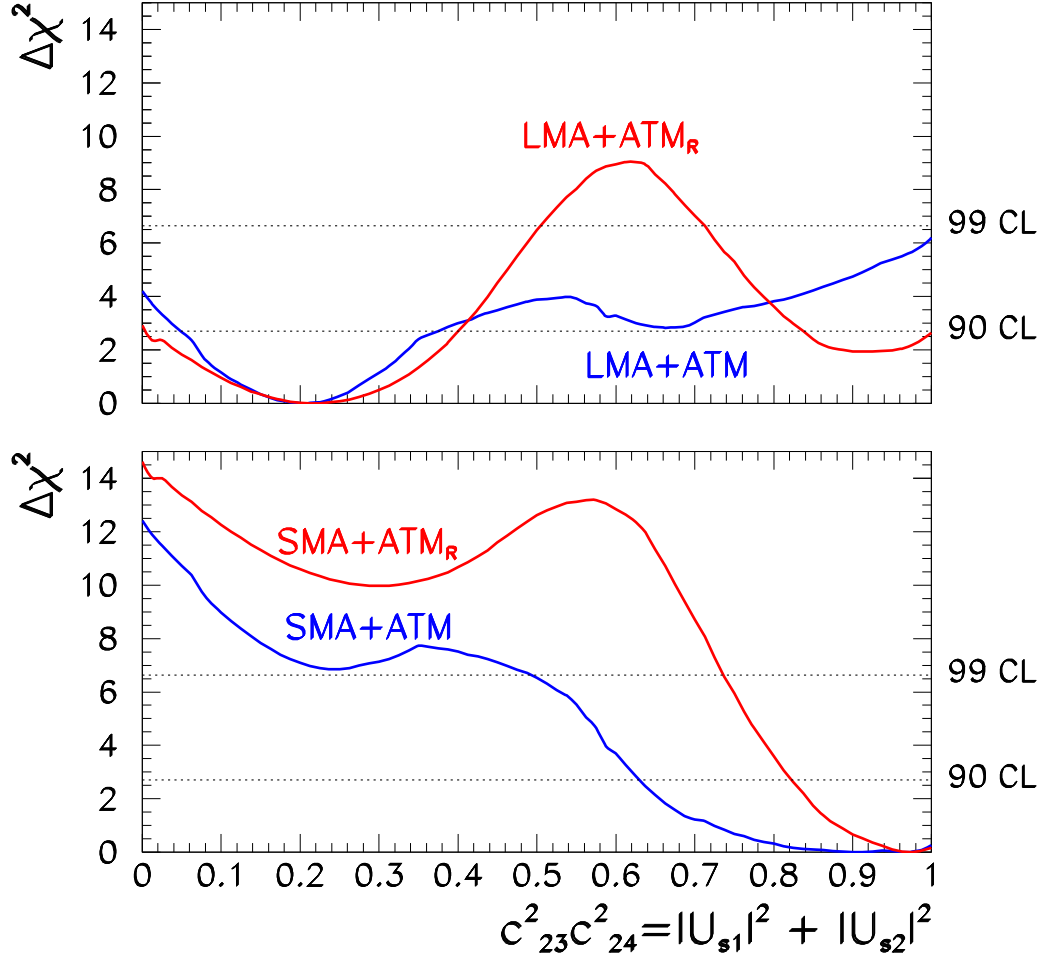,width=0.9\textwidth}}
    \end{center}
    \caption{$\Delta \chi^2$ as a function of the active--sterile admixture
      $|U_{s1}|^2 + |U_{s2}|^2 =c_{23}^2 c_{24}^2$. In the ATM curves $\chi^2$
      have been minimized with respect to $\vartheta_{34}$, $\Delta m^2_{43}$
      and the $\nu_\mu$ projection $|U_{\mu 1}|^2 + |U_{\mu 2}|^2 = s_{23}^2$,
      as well as with respect to the 1--2 parameters $\vartheta_{12}$ and
      $\Delta m^2_{21}$. The ATM$_{\rm R}$ curves correspond to the
      ``restricted'' case $|U_{\mu 1}|^2 + |U_{\mu 2}|^2 = s_{23}^2 = 0$, so
      $\chi^2$ has been minimized only with respect to $\theta_{34}$, $\Delta
      m^2_{43}$ and the 1--2 parameters. In the upper (lower)  
      panel the 1--2 parameters
      $\vartheta_{12}$ and $\Delta m^2_{21}$ have been constrained to lie in
      the LMA (SMA) solution region.}
    \label{chi2_combined}
\end{figure}

To address this question in full generality we have studied the behaviour of
the global $\chi^2_{\rm atm+sol}$ function defined as
\begin{eqnarray}
    \chi^2_{\rm atm+sol}(\Delta m^2_{21}, \vartheta_{12},\Delta m^2_{43}, 
    \vartheta_{34},\vartheta_{23},\vartheta_{24}) &=& \nonumber \\
    \chi^2_{\rm sol} (\Delta m^2_{21}, \vartheta_{23},c_{23}^2 c_{24}^2) &+&
    \chi^2_{\rm atm}(\Delta m_{43}^2,\vartheta_{34},c_{23}^2,c_{24}^2)
\end{eqnarray}
with the mixing angles $\vartheta_{23}$ and $\vartheta_{24}$ determining the
active--sterile admixture for the following scenarios:
\begin{itemize}
  \item ATM+SOL$_{\rm LMA (SMA)}$: in this case we minimize
    $\chi^2_{\rm atm+sol}$ with respect to $\Delta m^2_{21}$,
    $\vartheta_{12}$, $\Delta m^2_{43}$ and $\vartheta_{34}$ with $\Delta
    m^2_{21}$ and $\vartheta_{12}$ constrained to be in the LMA (SMA) region.
    The $\nu_\mu$ projection $s^2_{23} = |U_{\mu 1}|^2 + |U_{\mu 2}|^2$ is
    free to vary in the range allowed by the analysis.
  \item ATM$_{\rm R}$+SOL$_{\rm LMA(SMA)}$: in this case we fix
    $s^2_{23} = |U_{\mu 1}|^2 + |U_{\mu 2}|^2 = 0$, so $|U_{s 1}|^2 + |U_{s
    2}|^2 = c_{24}^2$. We minimize $\chi^2_{\rm atm+sol}$ with respect to
    $\Delta m^2_{21}$, $\vartheta_{12}$, $\Delta m^2_{43}$ and
    $\vartheta_{34}$, with $\Delta m^2_{21}$ and $\vartheta_{12}$ constrained
    to be in the LMA (SMA) region.
\end{itemize}
In Table~\ref{table_comb} we give the values $\chi^2_{\rm min, atm+sol}$ of
the best-fit point and its associated GOF in the six-dimensional
(five-dimensional for the ``restricted'' case) space in each of these cases
together with the best-fit value of the mixings $|U_{\mu 1}|^2 + |U_{\mu 2}|^2
= s^2_{23}$ and $|U_{s 1}|^2 + |U_{s 2}|^2 = c_{23}^2 c_{24}^2$. In
Fig.~\ref{chi2_combined} we plot the shift in $\chi^2$ from the corresponding
global minimum in each of the scenarios as the mixing $|U_{s 1}|^2 + |U_{s
2}|^2 = c_{23}^2 c_{24}^2$ varies. From the table and the figure we find the
following behaviours:
\begin{itemize}
  \item There are two favourite configurations, which we will denote as
    near-pure-sterile solar neutrino oscillations plus near-pure-active
    atmospheric neutrino oscillations (NPSS+NPAA), and close-to-active solar
    neutrino oscillations plus close-to-sterile atmospheric neutrino
    oscillations (CAS+CSA), respectively. NPSS+NPAA oscillations are
    characterized by the minima
    \begin{equation}
        |U_{s1}|^2 + |U_{s2}|^2 \sim 0.91 \mbox{--} 0.97 \;,
        \label{nps}
    \end{equation}
    where the exact position of the minimum depends on the allowed admixture
    $|U_{\mu1}|^2 + |U_{\mu 2}|^2$. CAS+CSA oscillations are characterized by
    the minima
    \begin{equation}
        |U_{s1}|^2 + |U_{s2}|^2 \sim  0.18 \mbox{--} 0.2 \;.
        \label{cpa}
    \end{equation}
    In both cases the minima are not very deep in $\chi^2$ (this is
    particularly the case for CAS+CSA).
  \item In none of the cases is the maximal active--sterile (MAS) admixture
    $|U_{s 1}|^2 + |U_{s 2}|^2 =c_{23}^2 c_{24}^2 = 0.5$ favoured.
\end{itemize}
The scenarios with the solar solution within the LMA region
(Fig.~\ref{chi2_combined}b) prefer the CAS+CSA scenario, although the
dependence of $\chi^2_{\rm atm+sol}$ with the active--sterile admixture is not
very strong and it presents a secondary minimum near the complementary
situation of close-to-sterile solar plus close-to-active atmospheric
(CSS+CAA) scenario, which is acceptable at 90\% CL. They give the best fit to
the combined analysis with a GOF: 67\%
for the ``unrestricted'' case. The ``restricted'' case gives a 
worse fit but still with an acceptable GOF of 59\%. 
In these scenarios MAS admixture
is ruled out at 99\% CL for the ``restricted'' case and it is acceptable at
about 95\% CL for the unrestricted one. This behaviour arises from the fact
that in these scenarios increasing the sterile admixture both in the
atmospheric pair and in the solar case is strongly disfavoured. It is
precisely in this case that one would expect the combined solution to lie near
MAS admixture. However, the results show that it is the solar dependence
(which dominates in the combination) that leads to the slightly better
description for the CAS+CSA scenario with a secondary minimum at the
complementary CSS+CAA scenario.

On the other hand, scenarios with the solar solution within the SMA region
(Fig.~\ref{chi2_combined}b) prefer the NPSS+NPAA scenario. They give 
the worse 
fit to the combined analysis for the ``unrestricted'' case with a GOF of 
62\% improving to 65\% for the ``restricted'' case . In these scenarios
the dependence of $\chi^2_{\rm atm+sol}$ with the active--sterile admixture is
more pronounced and it is dominated by the dependence of the atmospheric
neutrino analysis, although there is a secondary minimum at the CAS+CSA
configuration, which for the ```unrestricted'' case is acceptable at 99\% CL.
In all these scenarios MAS admixture is ruled out at more than 99\% CL.
Qualitatively this behaviour arises from the fact that, in these scenarios,
increasing the sterile admixture in the atmospheric pair is strongly
disfavoured while it is still acceptable for the solar pair since the SMA
solution is valid for pure $\nu_e\rightarrow\nu_s$ solar oscillations.

\acknowledgments
The work of M.M. is supported by the European Union
Marie-Curie fellowship HPMF-CT-2000-01008.
MCG-G is supported by the European Union Marie-Curie fellowship
HPMF-CT-2000-00516.
This work was also supported by the Spanish DGICYT under grants PB98-0693
and PB97-1261, by the Generalitat Valenciana under grant
GV99-3-1-01, by the  European Commission RTN network 
HPRN-CT-2000-00148 and by the European Science
Foundation network grant N.~86.

\newpage 
\begin{table}
    \catcode`?=\active \def?{\hphantom{0}}
    \begin{tabular}{|l|c|c|c|c|}
        Scenario &  $(|U_{\mu 1}|^2+|U_{\mu 2}|^2 = s_{23}^2)_{\rm min}$ &
        $(|U_{s1}|^2+|U_{s2}|^2 = c_{23}^2c_{24}^2)_{\rm min}$ &
        $\chi^2_{\rm min}$ & GOF \\
        \hline
        ATM+SOL$_{\rm LMA}$            & 0.065 & 0.21 & 73.8 & 67\% \\
        ATM$_{\rm R}$+SOL$_{\rm LMA}$  & 0.??? & 0.22 & 77.4 & 59\% \\
        ATM+SOL$_{\rm SMA}$            & 0.030 & 0.91 & 75.4 & 62\% \\
        ATM$_{\rm R}$+SOL$_{\rm SMA}$  & 0.??? & 0.98 & 75.5 & 65\% \\
    \end{tabular}
    \caption{$\chi^2_{\rm min, atm+sol}$ of the best-fit point and its
      associated GOF in the six-dimensional (five--dimensional for the
      ``restricted'' case) space in each of the cases discussed in the text
      together with the best fit value of the mixings $s^2_{23} = |U_{\mu
      1}|^2 + |U_{\mu 2}|^2$ and $c_{23}^2c_{24}^2 = |U_{s 1}|^2 + |U_{s
      2}|^2$.}
      \label{table_comb}
\end{table}

\end{document}